%% file: resilience.tex
\documentclass[a4paper]{article}

\usepackage{moreverb}

 \usepackage[pdftex,colorlinks,bookmarksopen,bookmarksnumbered,citecolor=red,urlcolor=red]{hyperref}
 \usepackage{graphicx}
 \usepackage{amsmath,amssymb}
 \usepackage{booktabs, rotating, tabularx, url}
 \usepackage{todonotes}
 \usepackage{algorithm}
 \usepackage{algpseudocode}
 \usepackage{multirow}
 \usepackage{color}
 \usepackage{algorithmicx}

\newcommand\BibTeX{{\rmfamily B\kern-.05em \textsc{i\kern-.025em b}\kern-.08em
T\kern-.1667em\lower.7ex\hbox{E}\kern-.125emX}}

\newcommand{\commentGeneric}[2]{~\\\fbox{\begin{minipage}{0.9\columnwidth}
 \textbf{Kommentar #1:}\\[1ex]
 \begin{minipage}{0.975\textwidth}
 \it #2
 \end{minipage}\end{minipage}}\quad\\[1ex]}
\newboolean{SHOWCOMMENTS}
\setboolean{SHOWCOMMENTS}{false}
\ifthenelse{\boolean{SHOWCOMMENTS}}{%
\newcommand{\commentG}[1]{\commentGeneric{Gmeiner}{#1}}
\newcommand{\commentH}[1]{\commentGeneric{Huber}{#1}}
\newcommand{\commentS}[1]{\commentGeneric{St\"urmer}{#1}}
\newcommand{\commentR}[1]{\commentGeneric{R\"ude}{ #1}}
\newcommand{\commentU}[1]{\commentGeneric{R\"ude}{#1}}
\newcommand{\commentW}[1]{\commentGeneric{Wohlmuth}{#1}}

}
{%
\def\commentG#1{}
\def\commentH#1{}
\def\commentS#1{}
\def\commentR#1{}
\def\commentW#1{}
\def\commentU#1{}
\def\todo#1{}
}

 \graphicspath{{Figure/}}

\begin{document}
% =============================================================================

%\runningheads{M.~Huber et al.}{Resilience for Multigrid}

\title{Resilience for Exascale Enabled Multigrid Methods}

\author{M.~Huber$^{\mbox{\footnotesize 1,}\ast}$, B.~Gmeiner$^{\mbox{\footnotesize 2}}$, 
U.~R\"ude$^{\mbox{\footnotesize 2}}$, B.~Wohlmuth$^{\mbox{\footnotesize 1}}$ }%\affilnum{1}\corrauth, B.~Gmeiner\affilnum{2}, U.~R\"ude\affilnum{2}, B.~Wohlmuth\affilnum{1}}
\date{}

%\address{\affilnum{1}Insitute for Numerical Mathematics (M2), Technische Universit\"at M\"unchen, %
%\affilnum{2} Department of Computer Science 10, FAU Erlangen-N\"urnberg}

%\corraddr{Journals Production Department, John Wiley \& Sons, Ltd,
% The Atrium, Southern Gate, Chichester, West Sussex, PO19~8SQ, UK.}
%\corraddr{huber@ma.tum.de}

\maketitle
{\footnotesize \centering $^{\mbox{\tiny 1}}$Institute for Numerical Mathematics (M2), Technische Universit{\"a}t M{\"u}nchen; 
$^{\mbox{\tiny 2}}$Department of Computer Science 10, FAU Erlangen-N\"urnberg; $^{\ast}$Corresponding author: \texttt{huber@ma.tum.de}}

\begin{abstract}
  \input{abstract}

\end{abstract}

% \keywords{}
\vspace{10pt}

 \commentH{Sollen wir hier nicht lieber Massive Parallelism rausnehmen?}

% =============================================================================
% INTRODUCTION
% =============================================================================
 \section{Introduction}\label{sec:intro}
 \input{introduction}

% ============================================================================
% Fault Situation
% ============================================================================
 \section{Faulty Solution Process}\label{sec:faultsolprocess}
 \input{faultsituation}

% ============================================================================
% Local Recovery Strategy
% ============================================================================
 \section{Local Recovery Strategy}\label{sec:localrecovery}
 \input{localrecoverystrat}

% =============================================================================
% CONCLUSION
% =============================================================================
\section{Conclusion and Outlook}\label{sec:conclusion}
\input{conclusion}

\clearpage
\bibliographystyle{abbrv}
\bibliography{resilience}

\end{document}

%% file: abstract.tex
%!TEX root = ./resilience.tex
With the increasing number of components and further miniaturization 
the mean time between faults in supercomputers will decrease.
System level fault tolerance techniques
are expensive and cost energy, since they are often based on redundancy.
Also classical check-point-restart techniques reach their limits when the time
for storing the system state to backup memory becomes excessive.
Therefore, algorithm-based fault tolerance mechanisms can become an attractive alternative.
This article investigates
the  solution process for elliptic 
partial differential equations  that are discretized by 
finite elements.  
Faults that occur in the parallel geometric multigrid solver
are studied in various model scenarios.
In a standard domain partitioning approach, 
the impact of a failure of a 
core or a node will affect one or several subdomains. 
Different strategies are developed to 
compensate the effect of such a failure algorithmically.
The recovery is achieved by solving a
local subproblem with Dirichlet boundary conditions using 
local multigrid cycling algorithms.
Additionally, we propose a superman strategy where extra  
compute power is employed to minimize the time of the recovery process.

\commentU{bin mir nicht sicher, dass wir es work balance oder work distribution nennen sollten. Und eigentlich sollte das abstract auch *noch* direkter auf den Kern hin geschrieben werden, am Anfang des Abstract ist die Hinfuehrung zu behaebig, bzw. wir sollten nicht herum reden sondern gleich mit dem Kern loslegen}
\commentH{Abstract noch nicht wirklich angefasst.}

%% file: introduction.tex
%!TEX root = ./resilience.tex

Future high performance systems will be characterized by millions of compute nodes that are executing up to a billion of parallel threads.
This compute power will be expensive due to the associated investment and operational cost which will include energy 
consumption as a rapidly increasing factor.
Hardware, software and algorithmic components of such a large scale  
computing are interdependent, and 
thus reliability of each single component 
is necessary.
Since this increasing complexity 
results in a higher probability of any kind of failure in the HPC-system \cite{Cappello09}, 
strategies which circumvent and/or accomplish such a behavior are inevitable. A commonly used approach is copying the data 
to backup disks on the I/O-system and restart the run with the stored data in case of a failure. These checkpoint/restart strategies need to
collect and transfer data from and to all processors and are, in general, too costly and not attractive.

Geometric multigrid methods 
can deliver an asymptotic optimal complexity and can be implemented with excellent efficiency on large scale parallel machines \cite{BergenHuelsemann04,GRSWW13,GKSR14,Sundar12}. 
Typical future runs involving multigrid computations will last from a few hours to weeks and use up to a billion threads. Therefore, an error-resilient methodology for
any failure will be required and necessary to obtain fault-free and efficiently computed solutions. 

The type of failures and their treatment can be categorized in hardware-based fault tolerance (HBFT) \cite{MKFM13,DJLMMP14,mukherjee2005,Malkowski10},
software-base fault tolerance (SBFT) \cite{BBBCD13,Zheng06,SBBC14,BDBHBD13}  and algorithm-based fault tolerance (ABFT) \cite{HA84,BFHH12,CSBS12}, 
 for a general overview of fault tolerance techniques we refer to \cite{Cappello09,CGGKS09,CGKKS14}. 

In this paper, we focus on the algorithmic approaches for handling failures. 
ABFT improves the reliability of the HPC-system in detecting the failure and correcting the results by implementing the resilience in the algoithm itself.
Originally, ABFT was proposed by Huang and Abraham \cite{HA84} for systolic arrays where due to checksums the persistency of the data involved 
in the algorithm is monitored and reconstructed. Later, it was extended to applications in linear algebra
such as addition, matrix operations, scalar product, LU-decomposition, transposition and in fast Fourier transformation 
\cite{AnfinsonLuk88,BoleyBrentGolubLuk92,ChenDongarra08,LukPark88}. Currently, the work by Davies and Chen \cite{DaviesChen13} efficiently deals with
fault detection and correction during the calculation for dense matrix operations. For iterative - linear and Krylov space - solvers for sparse matrix 
such as SOR, GMRES, CG-iterations the previous mentioned approaches are not suitable, since this can result in high overhead for sparse linear algebra
\cite{SKB12} and were consequently adapted by \cite{BFHH12,Chen2013,RoyBanerjee93,Stoyanov_Webster_2013}. Cui et al. propose in 
\cite{cui_error-resilient_2013} a technique to use the structure of a parallel subspace correction method such that the subspaces
are redundant on different processors and the workload is efficiently balanced. Further, an algebraic multigrid solver was analyzed in \cite{CSBS12} where the most vulnerable
components are identified and resilience techniques are explored.

\commentH{Unsere contribution steht weiter unten. Zu wenig?}
\commentH{Habs mal versucht umzuschreiben. Was haelst du davon?}
Here, we investigate in a fault tolerant parallel geometric multigrid method.
Similar to \cite{cui_error-resilient_2013} we pursue fault tolerant strategies which 
\begin{itemize}
 \item converge when a fault occurs assuming it is detect\-able,
 \item minimize the delay in the solution process,
 \item minimize computational and communication overhead.
\end{itemize}
In order to achieve these goals, we study the consequences of failures for a geometric multigrid algorithm from bottom up. 
By applying local correction methods, we recover a partial solution and use it such that the 
effect of the fault on the global solution procedure is minimized.
The major difference to other approaches in ABFT \cite{CSBS12,cui_error-resilient_2013} 
is that we proceed without check-pointing data of the solution but rather
recalculate the faulty part.

Our paper is organized as follows: In Sec.\ \ref{sec:faultsolprocess}, we describe the model equation and briefly discuss 
the parallel hierarchical hybrid multigrid (HHG) framework that will serve as the basis of the study in this paper. 
Next, we introduce the failure scenario that is used to study
the  influence of a fault within a parallel geometric multigrid method and its effect on the convergence.
In Sec. \ref{sec:localrecovery}, we then develop  
local recovery strategies and demonstrate by numerical experiments how these improve the recovery behavior after the fault.

%% file: faultsituation.tex
%!TEX root = ./resilience.tex
\subsection{Model problem and geometric multigrid}\label{sec:ModelProblem}

This paper considers, for simiplicity of notation, the Laplace equation with Dirichlet boundary conditions
\begin{equation}\label{eq:Poisson}
-\Delta u = 0 ~~\mbox{in } \Omega, \qquad u = g ~~ \mbox{on } \partial \Omega 
\end{equation}
as model problem for the design and analysis of a fault recovery algorithm.
Here,  $\Omega \subset \mathbb{R}^3$ is a bounded polyhedral domain.
\begin{figure}[ht]
  \centering
  \includegraphics[width=0.425\textwidth]{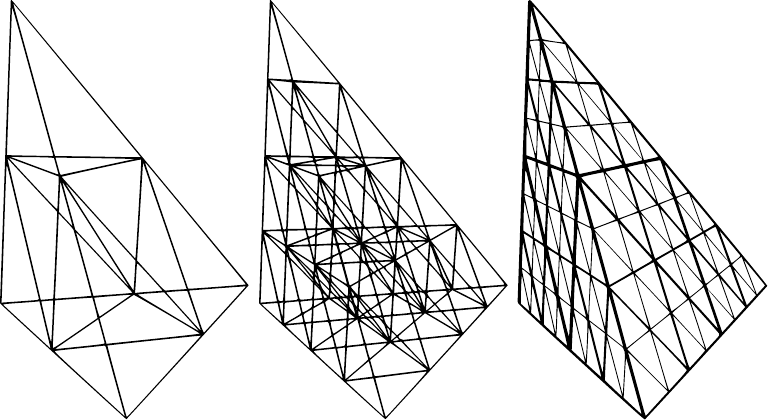}
  \caption{Structured tetrahedral refinement.}
  \label{fig:refinement}
\end{figure}
$\Omega$ is triangulated with an unstructured tetrahedral mesh
that we denote by $\mathcal{T}_{-2}$.
From this initial coarse mesh, a hierarchy of
meshes $\mathcal{T} := \{ \mathcal{T}_{l},~l=0, \ldots, L\}$ 
is constructed by successive uniform refinement
as illustrated in  Fig.~\ref{fig:refinement}. 

The discretization of \eqref{eq:Poisson}  
uses conforming linear finite elements (FE) on $\mathcal{T}_{l}$ 
that leads canonically to a nested sequence of finite 
element spaces $V_0 \subset V_1 \subset \ldots  \subset V_L \subset H^1(\Omega)$
and a corresponding family of  linear
systems 
\begin{equation}\label{eq:PoissonLin}
    A_l \underline{u}_l = \underline{f}_l;  ~ l=0, \ldots ,L.
\end{equation}
The Dirichlet boundary conditions are included in the linear systems \eqref{eq:PoissonLin}.

\commentH{Oder sollen wir $A_l u_l = 0$ schreiben und die BC sind punktweise erfuellt?}

This hierarchy will be used to set up an iterative  multigrid solver
and to define the error recovery strategy.
Multigrid methods can achieve level-inde\-pendent convergence rates
with optimal complexity $\mathcal{O}(N)$, where $N$ is the number of unknowns, cf. \cite{brandt2011multigrid,hackbusch1985multi}. 
We apply multigrid correction schemes in V-, \mbox{W-,} or F-cycles with standard components to \eqref{eq:PoissonLin}.
Explicitly, we use linear transfer operators and a hybrid variant of a Gauss-Seidel updating scheme as smoother.

\subsection{Hierarchical Hybrid Grids} \label{sec:HHG}
The Hierarchical Hybrid Grids (HHG) framework \cite{BergenHuelsemann04,GRSWW13,GKSR14}
is designed to combine the flexibility of unstructured FE meshes with the performance advantage of structured
grids in a block-structured approach.

The implementation is based on domain partitioning that splits the mesh into primitives:
vertices, edges, faces, and volumes. In the HHG data structure each primitive is then refined regularly resulting in a %
global block-structured grid.
For our later error recovery strategies, the domain partitioning approach is crucial, but the block-structured 
mesh structure could be generalized to fully unstructured meshes.
The multigrid operations such as smoothing, prolongation, restriction, and residual calculation,
are exploited such that they typically operate on the primitive itself and its neighboring primitives via ghost layers. 
These operations are inherently local and suited for parallel computations on a distributed memory system using
message passing with MPI. 
Here, the primitives are mapped to processors that execute the local operations.
The data dependencies require a systematic exchange of the ghost layers.
This functionality is provided in a transparent and highly optimized 
form in the HHG framework.

\subsection{Fault Model}\label{sec:FaultModel}
% \commentU{Ich habe kuerzlich gelernt, dass die Art der Faults, die wir ansehen
% "fail-stop" heisst. Es kann sein, dass wir darueber und Fragen, wie man solchen einen fail-Stop erkennt, noch etwas mehr schrieben muessen ... kann sein, dass wir hier Hilfe von einem Systemexperten (-> Wellein?) nuetzen sollten. Bjoern-Markus koennt ihr bei uns dazu mal unsere Kollegen Markus Stuermer und eventuell Sebastian Kuckuk befragen ob das reicht was wir haben. Wir muessen zumindest eine Ref auf fault-Tolerant MPI in das Paper machen (und nicht nur footnotes und URLs. 
% Wir muessen dann im Kontext damit schreiben, wie wir uns dessen Nutzung vorstellen
% und welche Funktionalitaet wir erwarten. 
% Dabei muessen wir klar sagen, was fehlt.
% Der Text unten ist dazu noch zu vage.
% Dieser technische Teil sollte moeglicherweise auch am Anfang des Abschnitts stehen, eventuell
% sogar einen eigenen Abschnitt bekommen.}

We assume that a failure in the solution process for \eqref{eq:PoissonLin} can occur at any time.
For our study, we concentrate on a specific fault model
under assumptions similar to \cite{cui_error-resilient_2013,Harding14}. 
We restrict the analysis, for simplicity, to the case that only one process crashes.
All % developed
strategies can be % easily 
extended easily to a defect of more processors, since
they only rely on the locality of the fault.

Furthermore, we concentrate on the case of using V-cycles for the solution of \eqref{eq:PoissonLin}.
The input tetrahedral mesh $\mathcal{T}_{-2}$ defines the partitioning
used for parallelization in HHG.
Each tetrahedron in $\mathcal{T}_{-2}$ is mapped to a processor, including all the refined subdomain meshes contained in the coarsest level element.
Consequently, the number of subdomains and the number of processes is equal to the 
number of tetrahedra in $\mathcal{T}_{-2}$.

If a process experiences a fault, the information of the subdomain is lost.
In the  context of this article, the {\em faulty} subdomain
$\Omega_F \subset \Omega$ is just a single tetrahedron in $\mathcal{T}_{-2}$.
The other tetrahedra constitute the {\em healthy} subdomain
$\Omega_H$, i.e. those tetrahedra that are not affected by the fault.
Healthy and faulty regions are separated by an interface
$\Gamma_I := \partial \Omega_H \cap \partial \Omega_F$.
In the finite element mesh, as implemented in HHG, the interface region
$\Gamma_I$ contains the nodes living on faces, edges, and vertices of the input mesh.
These data structures are responsible for handling communication in HHG and are thus stored redundantly
in the form of ghost instances on several processors. 
Thus, even if one of the instances is lost due to the fault, a complete recovery is always
possible for them, and thus we assume implicitly that the data associated with them are unaffected by the fault.

\begin{figure}[ht]
  \vspace{-0.3cm}
  \centering
  \includegraphics[width=.45\linewidth]{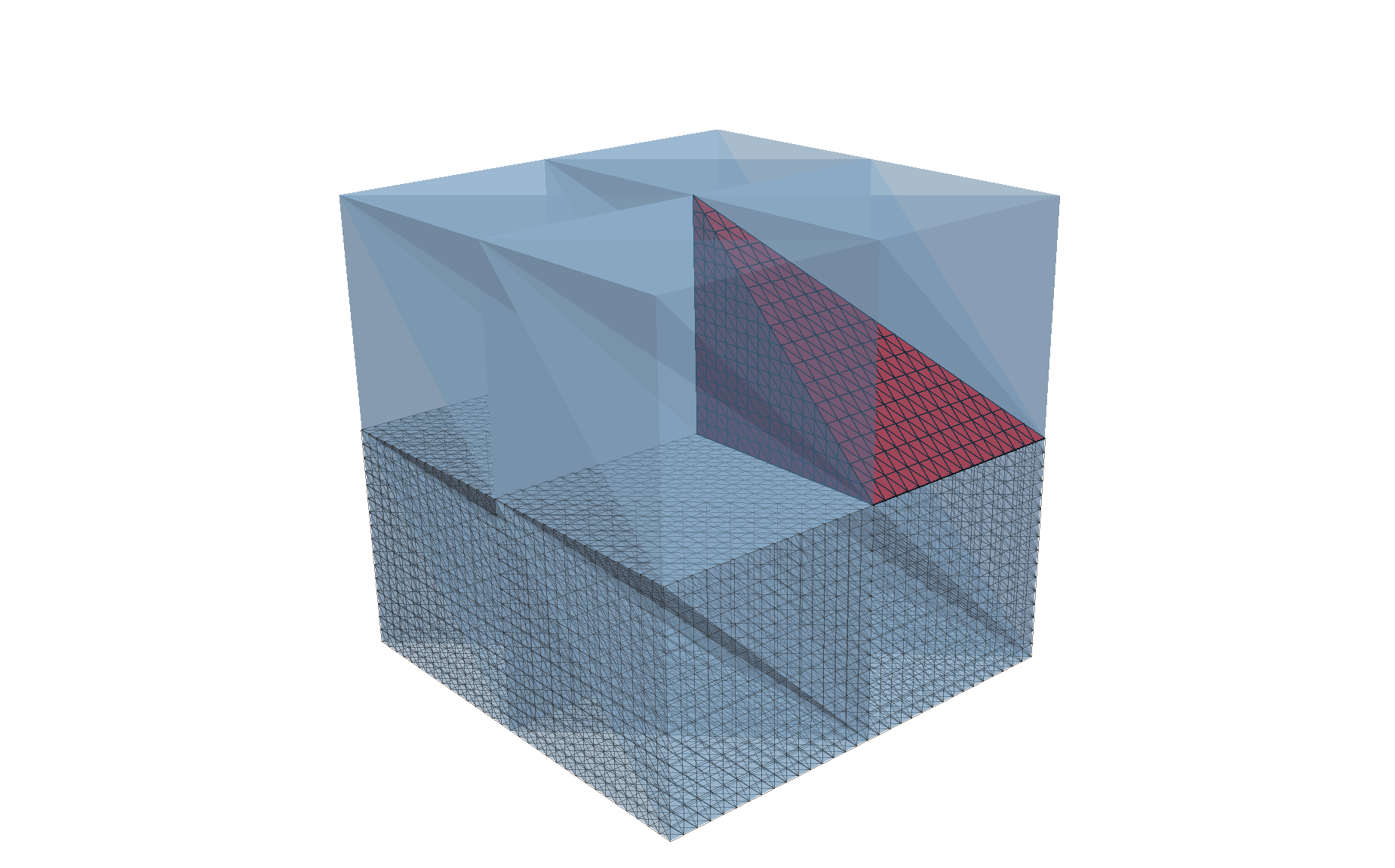}\hspace{-1cm}
  \includegraphics[width=.35\linewidth]{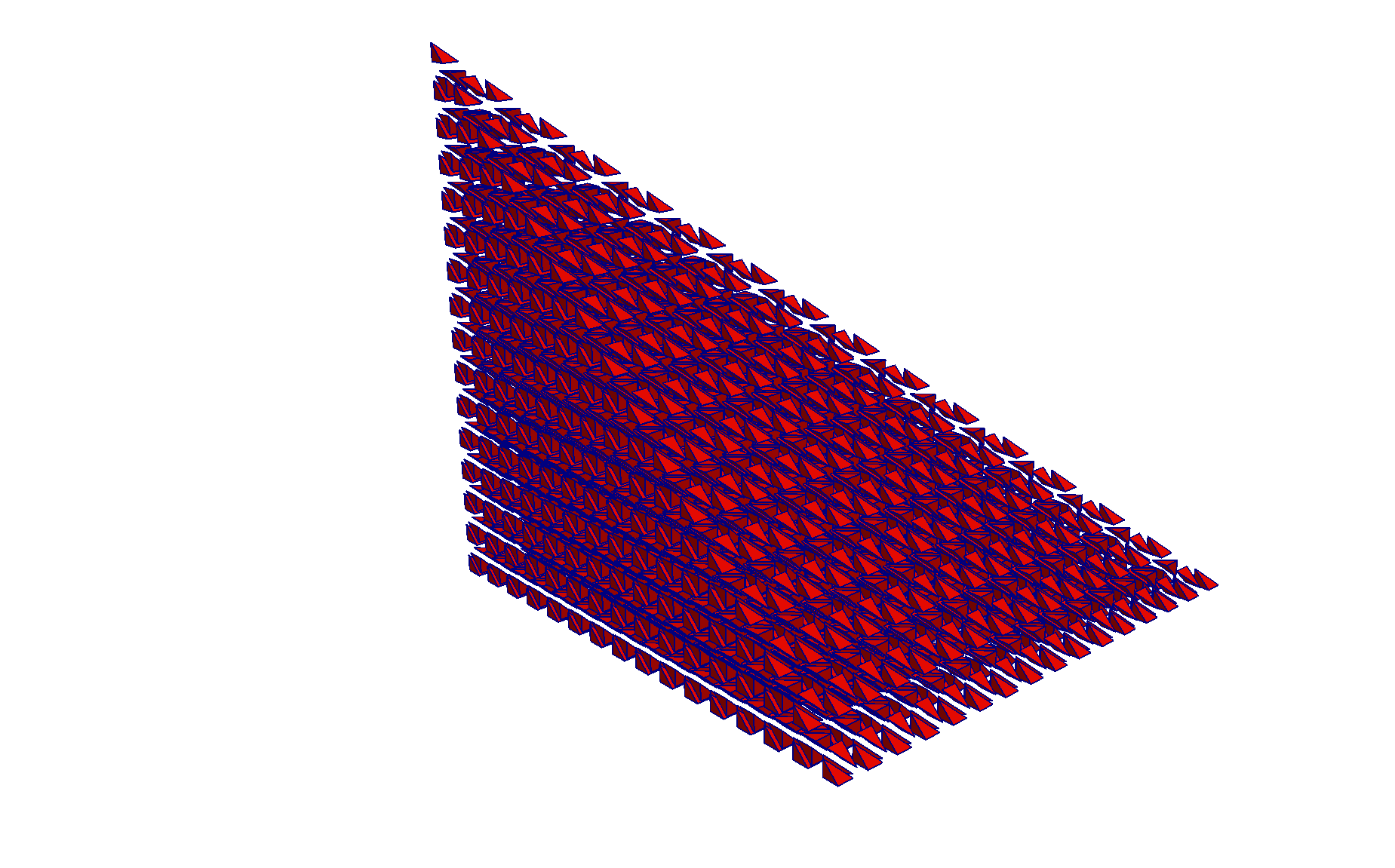}
  \caption{Fault of one process. Left: Computational domain (here: 16 mil. unknowns) with a faulty (red) subdomain, one input grid tetrahedron (here:
  300 000 unknowns). Right: Faulty subdomain.}
  \label{fig:FaultTet}
\end{figure}

In Fig. \ref{fig:FaultTet}, the setup is illustrated for a computational domain with 16 million unknowns.
The domain consists of  48 tetrahedral subdomains that are 
distributed to 48 processors. 
Then each subdomain includes 300 000 unknowns and, thus, the failure of a process causes the loss of information 
for 300 000 unknowns.

% Detecting and replacing an erroneous process,
% one might expect from a fault tolerant MPI solutions such as , although these 
% provided tools cannot support an instant reporting and replacing of processes as required for our simulations.
For our strategy, it is necessary that we can % adapt and 
detect erroneous processes quickly and then adapt the solution procedure dynamically.
Unfortunately, the current supercomputer systems and the fault tolerant MPI-extensions
such as Harness FT-MPI  \footnote{http://icl.cs.utk.edu/ftmpi/} or ULFM \footnote{http://fault-tolerance.org/} % specifications
do not yet support this functionality as ideally needed.
\commentH{Lassen wir das jetzt so mit dem FT-MPI?}
\commentU{Habs noch mal leicht umformuliert}
For the purposes of this study, we suppose a failure is reported as soon % and as fast
as it occurs during a multigrid cycle.
When a process crashes, we assign a new  - until then not used - 
substitute process to take over its job.
This assumes that a large scale parallel computation is started with
a certain number of substitute processors initially being idle -- very much like substitute players
in a team sport match.
The solution values % of the unknowns of 
in the faulty subdomain are set to zero as initial guess.
Other initial guesses could also be used, such as
data from previous check-pointing % data
or values obtained by interpolation,
but this will not be considered here in detail.
After the local re-initialization of the problem, 
we continue with multigrid cycles in the solution process.
\begin{figure}[ht]
 \includegraphics[width = 0.4\textwidth]{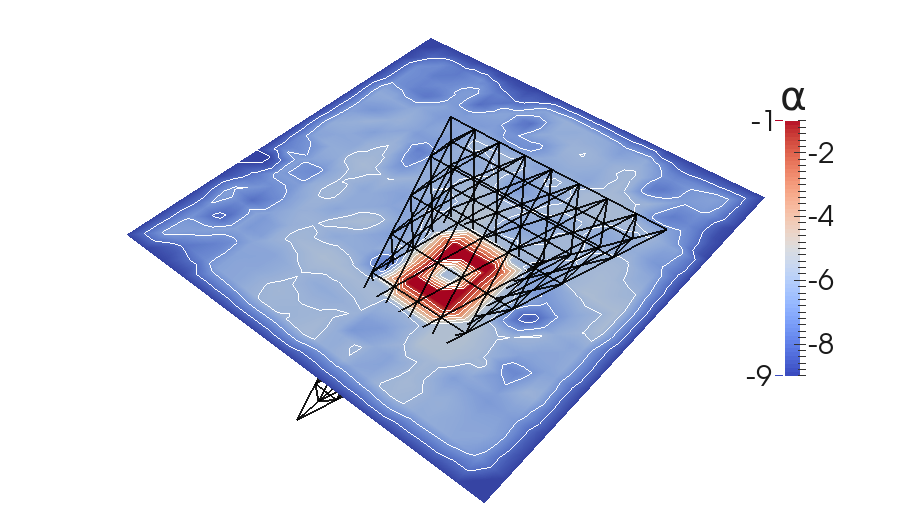}\hspace{1cm}
 \includegraphics[width = 0.4\textwidth]{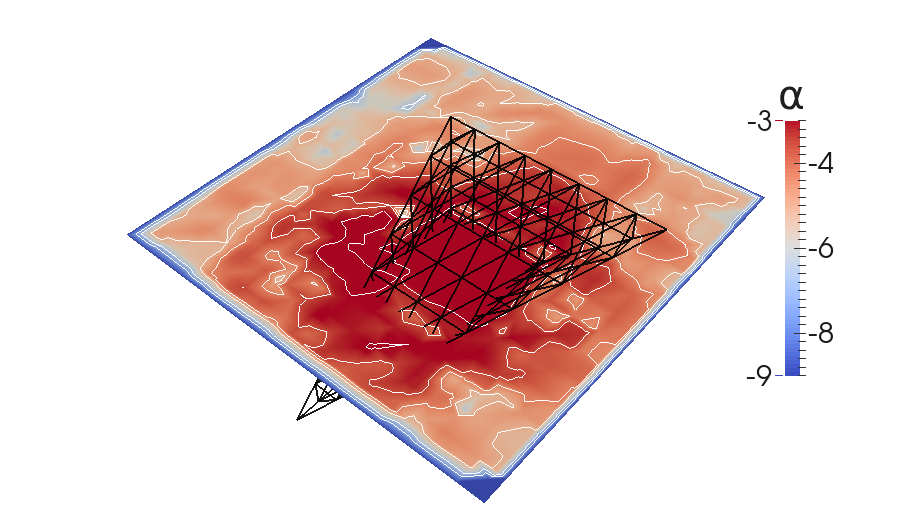}
 \caption{Cross section through the domain $\Omega$ and the surface of the faulty tetrahedron. Left figure:
  residual error directly after the failure.
Right figure: residual error after one additional global V-cycle. $\alpha := \log_{10}(|Residual|)$.}
 \label{fig:ResPlot}
\end{figure}

In a first experiment, we consider the performance of our multigrid iteration
when it is continued after a fault has occurred.
All numerical experiments are performed within the HHG framework % presented 
introduced in Subsec. \ref{sec:HHG}.
We choose the computational domain $\Omega = (0,1)^3$, and $ g = \sin(\pi (x+\sqrt{2}y)) \sinh(\sqrt{3} \pi z)$ in \eqref{eq:Poisson} with the described
setup in Fig. \ref{fig:FaultTet}.
In the solution process, we apply V-cycles with three pre- and post-smoothing steps of the Gauss-Seidel smoother of Sec. \ref{sec:ModelProblem}.

In Fig. \ref{fig:ResPlot}, the residual is visualized on a cross section
through the domain together with the surface of the tetrahedron 
where the fault had occurred after 5 global iterations.
Right after the failure and after re-initialization, the largest residual error is clearly located in this tetrahedron.
These large local error components are transported over the whole domain in the course of the following multigrid iterations.
Though each application of the 
smoother transports information only across a few neighboring mesh cells,
multigrid employs coarser grids recursively. 
The smoothing on these grids
leads to the global data exchange that is essential for the level-independent
convergence of multigrid iterations.
Therefore, though the residual is reduced efficiently (in the $L_2$-norm) by such iterations,
we observe a pollution of the residual error across the whole domain.
This 
can be seen in Fig.\ \ref{fig:ResPlot} at the right, where the overall residual 
error has been reduced after the additional V-cycle (note the scaling), but the
error now pollutes the whole domain.

\begin{figure}[ht]
  \vspace{-0.9cm}
  \centering
  \includegraphics[width = 0.6\textwidth]{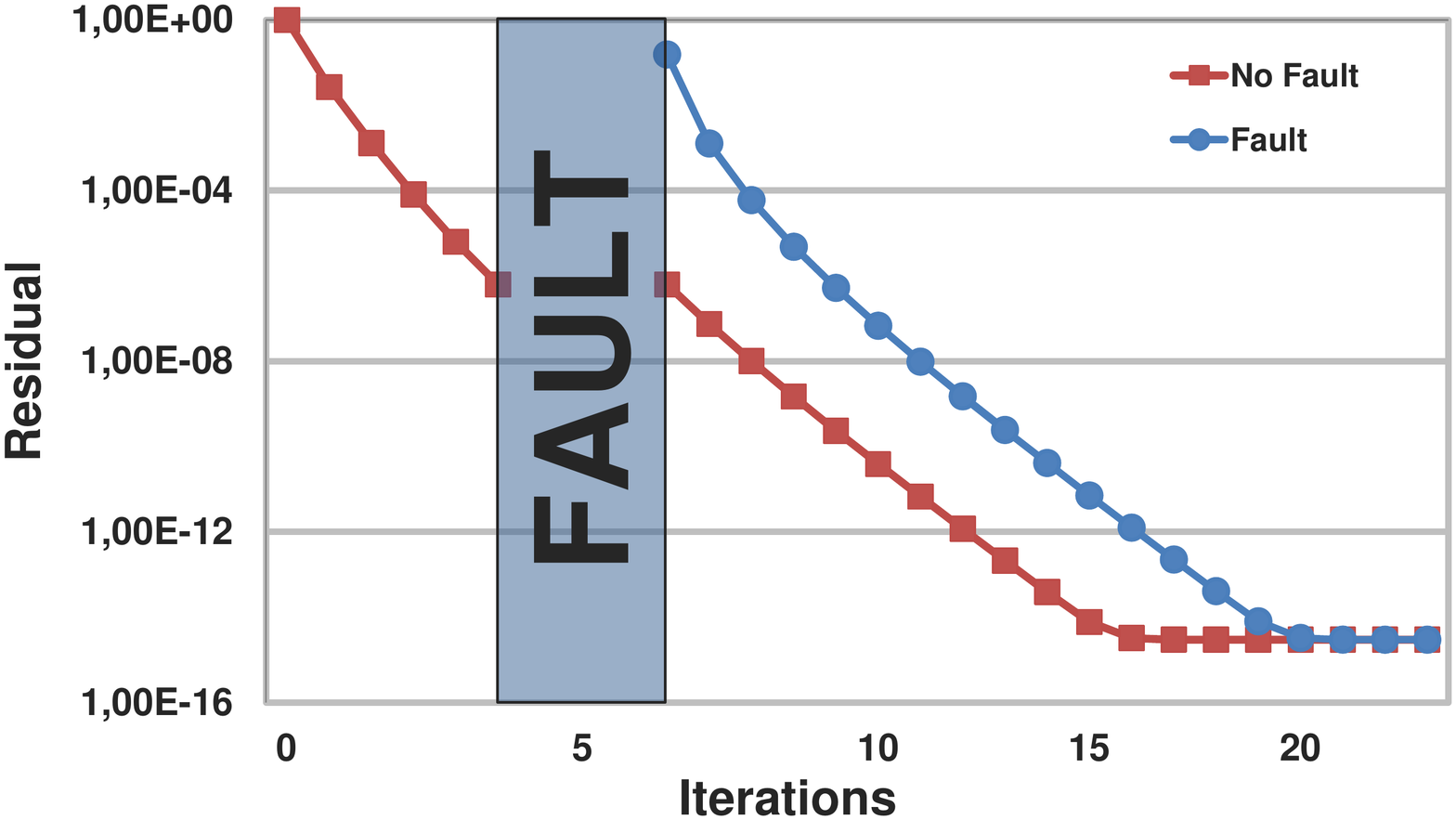}
  \vspace{-0.5cm}
  \caption{Convergence of the residual error scaled by the initial residual with fault after 5 iterations.}
  \label{fig:ResConv}
\end{figure}

\begin{figure}[ht]
  \centering
  \includegraphics[width = 0.98\textwidth]{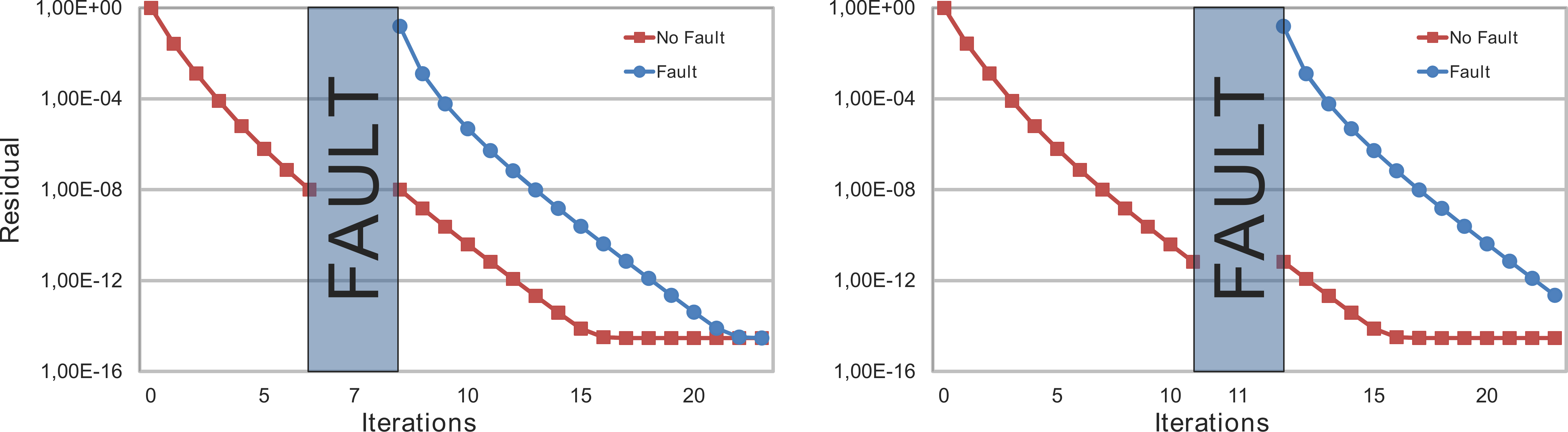}
  \caption{Convergence of the residual error scaled by the initial residual with fault after 7 iterations (left) and 11 iterations (right).}
  \label{fig:ResConv7_11}
\end{figure}

The numerical behavior is analysed quantitatively in Fig. \ref{fig:ResConv}.
After the fault the residual norm jumps from $6.24\cdot10^{-7}$ up to $1.55\cdot10^{-1}$. 
If a 
\textit{complete checkpointing-recovery} (CCR)
of the lost values could be performed,
it would fully restore the residual 
from before the fault.
Note, that this recovery, as marked in the diagram with
{\em no fault}, introduces
no additional computational effort
in comparison to the situation without failure, but 
writing and reading checkpoint data would be too expensive for large 
scale computations.
However, the failure introduces error components
that can be reduced efficiently by the multigrid method
as can be seen in the residuals marked with {\em fault}.
In the first cycles after the fault, we observe
a pre-asymptotic convergence rate that is better than the 
asymptotic rate for roughly three cycles.
This helps significantly to compensate for the fault.
The roundoff error limit of approximately 
$10^{-15}$ is reached after a total of 20 V-cycles, as compared to 16 V-cycles that were necessary in the unperturbed computation.
As expected these effects can be seen more drastically,
when the fault occurs at a later step during the iteration process. 
The situation of a fault after 7 iterations is displayed 
in Fig. \ref{fig:ResConv7_11} (left) and after 11 iterations in Fig. \ref{fig:ResConv7_11} (right).
In those cases, the global residual is already quite small when the fault occurs,
and we need 7 and 10 more iterations, respectively, to obtain the rounding error limit of $10^{-15}$. 
%
%

%% file: localrecoverystrat.tex
%!TEX root = ./resilience.tex
%
Avoiding the global pollution observed in Subsec.\ \ref{sec:FaultModel} motivates a recovery strategy. 
For the recovery step, we here require that it is local, 
i.e., can be performed without communication. % effort.
For the overall efficiency of the recovery strategy, 
it is essential that it can be performed as quickly as possible.
We therefore propose a {\em superman strategy},
which simply means that
more resources are devoted to perform the recovery
in an attempt to reduce the recovery time.
This becomes especially attractive, since the 
recovery % concept 
procedure is local,
i.e.\ the {\em superman substitute processor} can focus 
its attention on a single subdomain.
Note, that during the local recovery,
the global solution process cannot continue
unaffected, since this would require to use values from the  
faulty subdomain.
This access to the faulty subdomain % has to 
must be prevented, since otherwise a global error pollution would occur.
The values in the faulty subdomain will only become available 
again, when the local recovery has been
completed. 

Technically the speedup of the superman strategy
can be accomplished by additional parallelization.
We propose here, that e.g.\ a full (shared memory) compute node is
assigned to perform the local
recovery for a domain that was previously handled by a single core.
This can be accomplished by a dedicated OpenMP parallelization for the recovery process.
Otherwise, of course, a further domain partitioning of the faulty region might be
used together with an additional local distributed memory parallelization.

We denote the speedup factor that is achieved with a  superman strategy for the
local recovery by $\eta_{\mbox{\small speedup}} \in [1, \infty)$, i.e., if 
$\eta_{\mbox{\small speedup}} = 1$, there is no speedup. 
For the case $\eta_{\mbox{\small speedup}} \rightarrow  \infty$, 
the recovery would cost no time.
For the moment, let us assume such a {\em perfect superman}
and that the local recovery step is in this sense free.
Let us define the following local subproblem 
\begin{equation}\label{eq:localPoisson}
-\Delta u = 0 \qquad \mbox{in } \Omega_F, %\qquad u = g ~~ \mbox{on } \partial \Omega_L %.
\end{equation}
with Dirichlet boundary conditions on $\Gamma_I$.

\commentH{Warum ist hier nicht $\Gamma_I = \partial \Omega_F$. $\Gamma_I$ kann doch auch Randdaten von $\Omega$ haben?}

Under the assumptions of Subsec.\ \ref{sec:FaultModel},
we set the lost values in the faulty region $\Omega_F \subset \Omega$ to zero and before continuing with global problem
\eqref{eq:Poisson} we solve \eqref{eq:localPoisson}.

The subproblem \eqref{eq:localPoisson} can in principle 
be solved by any method, e.g., the relaxation that is used as multigrid smoothing,
a direct solver, Krylov space iterations, or multigrid cycles. We denote  by $k_F$ the number of local
solver iterations.
After solving the subproblem \eqref{eq:localPoisson} with sufficient accuracy,
the solution process for the global problem \eqref{eq:Poisson} can be resumed.
The procedure is presented in Alg. \ref{alg:LocalAlg}. 

\begin{algorithm}
\caption{Local recovery algorithm}\label{alg:LocalAlg}
\begin{algorithmic}[1]
    \State Solve \eqref{eq:Poisson} by multigrid cycles.
    \If{Fault has occurred}
    \State \textbf{STOP} solving \eqref{eq:Poisson}.
    \State Recover boundary values $\Gamma_I$ by neighboring subdomains.
    \State Set values of the defect subdomain $\Omega_F$ to zero.
    \State Solve \eqref{eq:localPoisson} in $\Omega_F$ subject to Dirichlet boundary conditions on $\Gamma_I$ using
    \State $k_F$ local solver iterations  with speedup factor $\eta_{\mbox{\small speedup}}$.
    \State \textbf{RETURN} to step 1 with new computed values in $\Omega_F$. 
    \EndIf
  \end{algorithmic}
\end{algorithm}

To quantify the  number of multigrid cycles that are saved by using the local recovery strategy,
we introduce a reference parameter {\em Cycle Advantage} which we denote by $\kappa$.
We assume a process experiences a fault after $k$ cycles.
Then, we evaluate the residual after $K$ cycles with $k \leq K$. We choose $K$ such that
the stopping criteria is fulfilled in the no-fault case.
Here, $r^{(k_F)}_{K}$ and $r^{(0)}_{K}$ denote the residual
after a local recovery with $k_F$ iterations and when no-local-recovery ($k_F=0$) has been performed, respectively.

Then, it holds
\begin{equation} \label{eq:localres_globalres}
 \| r^{(k_F)}_{K}\| \leq \| r^{(0)}_{K} \| ,
\end{equation}
since the local strategy improves the residual error. 
In the case of no-local-recovery,
we need to apply additionally $\kappa$ cycles such that
\begin{equation}\label{eq:localres_is_globalres}
 \| r^{(k_F)}_{K} \| = \| r^{(0)}_{K+\kappa} \| = \mu^{\kappa} \cdot \|r^{(0)}_{K}\|,
\end{equation}
where $\mu$ is the convergence rate of one multigrid cycle. Thus, we solve for $\kappa$ in \eqref{eq:localres_is_globalres}
\begin{equation}\label{eq:AiC}
 \kappa = \log( \| r^{(k_F)}_{K} \|/\|r^{(0)}_K\| ) / \log(\mu).
\end{equation}
Note, that for the calculation of the reference parameter $\kappa$ we need two versions of computation runs, one with a fault and a local recovery
strategy and one with a fault and without a local recovery strategy.
A local recovery strategy with $\kappa = 0$ does not improve the residual error in comparison to no recovery strategy, whereas a higher $\kappa$ implies 
an improvement of the residual error of magnitude $\kappa$ in multigrid cycles. For example, if $\kappa = 5$, then, the 
residual error without a local recovery needs 5 additional cycles to achieve the same residual error as the local recovery strategy.

Let us consider again the example of Subsec. \ref{sec:FaultModel}. We  study five different local recovery strategies: 
Gauss-Seidel smoothing (Smth), Jacobi preconditioned CG (PCG) iterations, local V-cycles, local W-cycles and local F-cycles.

\begin{figure}[ht]\centering
 \hspace{-26pt}
 \includegraphics[width = 0.37\textwidth]{Fault6Pic.png}\hspace{-13.5pt}
  \includegraphics[width = 0.37\textwidth]{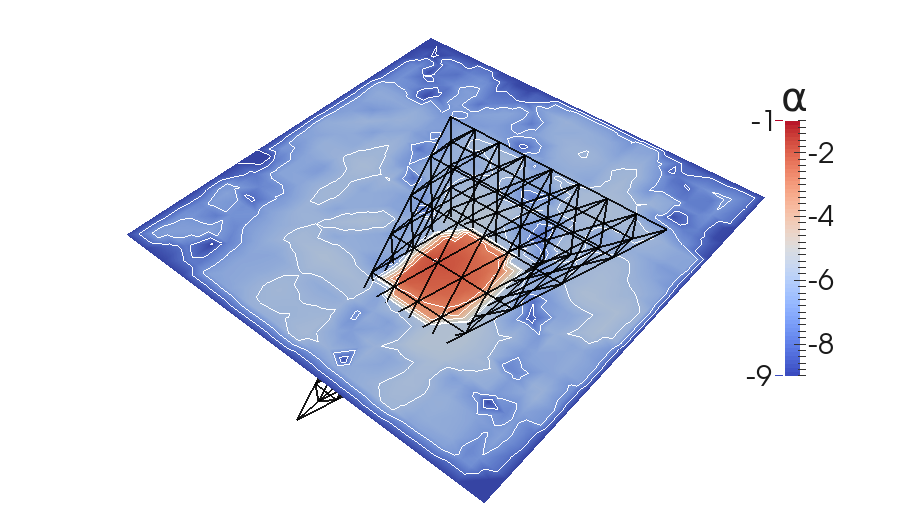}\hspace{-13.5pt}
 \includegraphics[width = 0.37\textwidth]{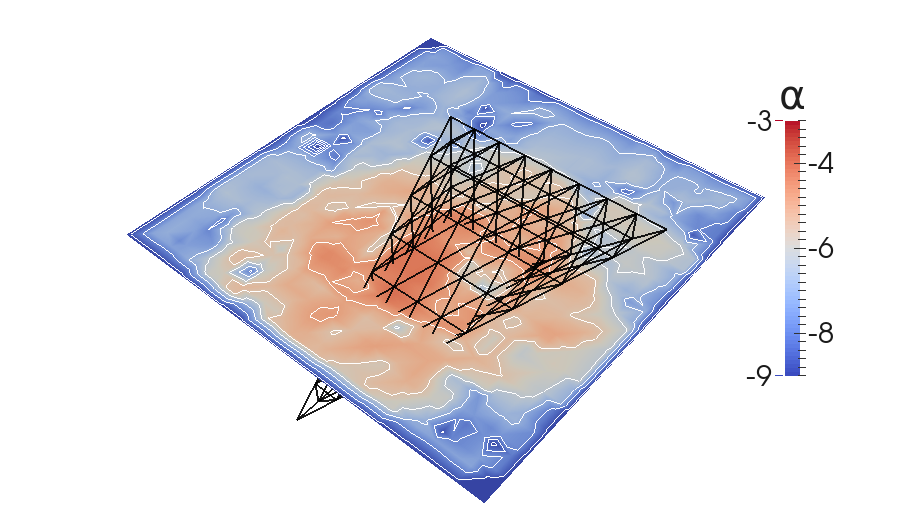}
 \caption{Cross section through the domain $\Omega$ and the surface of the erroneous tetrahedron. Left plot:
  residual error directly after the failure. Middle plot: residual error after applying a local recovery strategy (one local F-cycle). 
  Right plot: residual error after an additional V-cycle. $\alpha := \log_{10}(|Residual|)$.}
 \label{fig:PlotLocRec}
\end{figure}

In Fig. \ref{fig:PlotLocRec}, we present three cross sections through the computational domain together with the surface of the faulty tetrahedron. 
In the left plot, the residual error directly after the failure is shown, in the middle
plot the residual error after the local recovery with one F-cycle and in the right plot after an additional global V-cycle after the local recovery. 
We observe two major advantages of local recovery strategies compared 
to no-local-recovery: 
the local recovery reduces the residual error in the defective tetrahedron % primitive 
(middle plot of Fig. \ref{fig:PlotLocRec}) and the error pollution is much smaller over the computational domain (right plot of Fig. \ref{fig:PlotLocRec}).

\begin{figure}[t]
\centering
\includegraphics[width=0.6\textwidth]{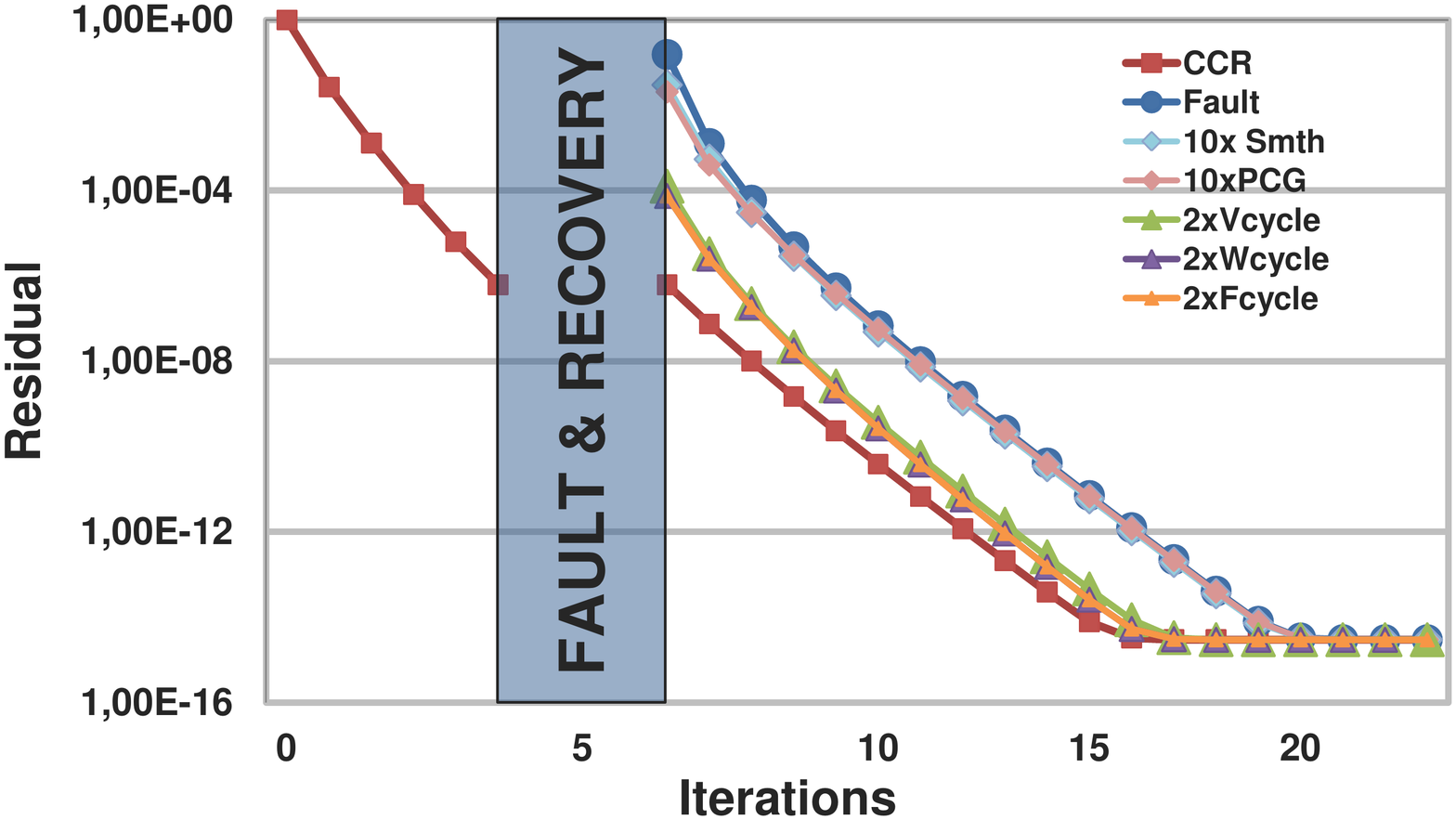}
\vspace{-0.5cm}
\caption{Convergence of the residual error scaled by the initial residual for different local recovery strategies.}
\label{fig:ResiConvLocDiff}
\end{figure}

\begin{figure}[t]
\vspace{-1.0cm}
\centering
\includegraphics[width=0.6\textwidth]{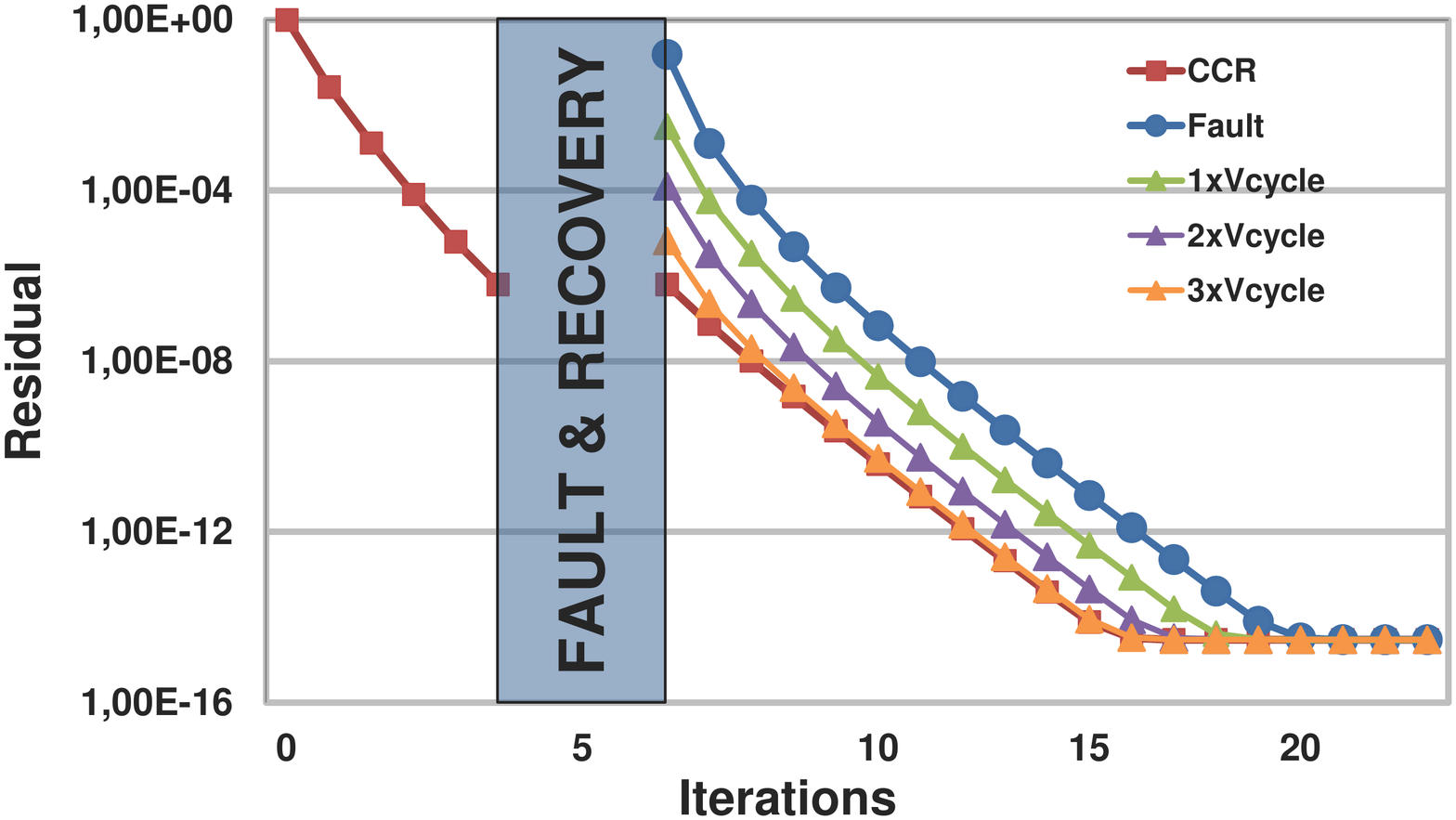}
\vspace{-0.7cm}
\caption{Convergence of the residual error scaled by the initial residual for different number $k_F$ of local V-cycles.}
\label{fig:ResiConvLocVcycle}
%\vspace{-1cm}
\end{figure}

In Fig. \ref{fig:ResiConvLocDiff}, we consider different local recovery strategies and their impact on the solution process. We observe that all
local strategies improve the residual error directly after the fault in comparison to using 
no-local-recovery. The different cycles, local V- (green), 
local W- (purple), and local F-cycle (orange) strategies, produce almost the same effect on the residual errors. 
However, using 10 PCG-iterations (pink) or 10 Gauss-Seidel 
smoothing steps (light blue) result only in an almost negligible improvement
compared to using no recovery at all.
After the fault and recovery, all recovery strategies using a local multigrid solver exhibit
a favorable pre-asymptotic convergence in the first three global iterations after recovery
and, then, align with the convergence of CCR. The residual error of 
the smoothing or PCG strategy after three further iterations is similar to the approach without a local strategy and, therefore, needs as many iterations as 
the no-local-recovery strategy to reach the prescribed accuracy of $10^{-15}$, i.e., three additional iterations in comparison to 
the CCR strategy. Two local V-cycles improve the situation such that the delay in finding the solution is reduced to one iteration. 
A local correction by two W-cycles or F-cycles deliver almost the same residual error and reduces the delay
in comparison to the two local V-cycles strategy. 
A F-cycle is preferred, since its computational cost is by a factor of $8/7$ lower than for a W-cycle. Further in Fig. \ref{fig:ResiConvLocVcycle},
we vary the number of V-cycles $k_F$ for solving the subproblem on $\Omega_F$. The delay in finding the solution significantly depends on how accurate
the subproblem is solved. One local V-cycle reduces the solution process by one iteration in comparison to no-local-recovery strategy,
two V-cycles by three iterations and three V-cycles completely compensate the fault.
Additionally, we compare the size of $\kappa$
for different local recovery strategies and for different iterations after which a failure
occurs. We study faults after 5, 7, 11 iterations, respectively, and smoothing, PCG-iterations, and
multigrid V-, W-, and F-cycle as local recovery strategies. The V-cycle convergence rate has been numerically evaluated as $\mu = 0.194$.
The cycle advantage $\kappa$ is measured after 15 iterations ($K=15$) and presented in Tab. \ref{tab:DiffFaults}. 

\begin{table}[ht]
  \footnotesize\centering
  \caption[Tab.]{Cycle Advantage for an early (after 5 iterations) , middle (after 7 iterations) , and late (after 11 iterations) 
		fault for different local recovery strategies.}
\vspace{0.05\textwidth}
  \label{tab:DiffFaults}

 \begin{tabular}{@{}*{1}{p{60pt}}@{}@{}*{1}{p{30pt}}@{}}
  \toprule
  \multicolumn{2}{c}{\textbf{Fault After 5 Iter.}}\\
  \cmidrule{1-2}
  ~Strategies~~~  &  ~~~~$\kappa$~~~ \\
  \cmidrule(r){1-1}\cmidrule(r){2-2} 
   CCR			&	4.164		\\
  1 $\times$ Vcycle	&	1.637		\\
  2 $\times$ Vcycle	&	3.075		\\
  3 $\times$ Vcycle	&	4.069		\\
  4 $\times$ Vcycle	&	4.165		\\
  5 $\times$ Vcycle	&	4.165		\\
  \cmidrule(r){1-1}\cmidrule(r){2-2}
  1 $\times$ Wcycle	&	1.825		\\
  2 $\times$ Wcycle	&	3.425 		\\
  3 $\times$ Wcycle	&	4.147		\\
  4 $\times$ Wcycle	&	4.164		\\
  5 $\times$ Wcycle	&	4.165		\\
  \cmidrule(r){1-1}\cmidrule(r){2-2}
  1 $\times$ Fcycle	&	1.828		\\
  2 $\times$ Fcycle	&	3.426		\\
  3 $\times$ Fcycle	&	4.147		\\
  4 $\times$ Fcycle	&	4.164		\\
  5 $\times$ Fcycle	&	4.165		\\
  \cmidrule(r){1-1}\cmidrule(r){2-2}
  2 $\times$ PCG 	&	0.007		\\
  5 $\times$ PCG 	&	0.026		\\
  10 $\times$ PCG 	&	0.037		\\
  \cmidrule(r){1-1}\cmidrule(r){2-2}
  10 $\times$ Smth	&	0.105		\\
  20 $\times$ Smth	&	0.172		\\
  30 $\times$ Smth	&	0.221		\\
  40 $\times$ Smth	&	0.260		\\
  50 $\times$ Smth	&	0.291		\\
  \bottomrule
 \end{tabular}
 \hspace{5pt}
 \begin{tabular}{@{}*{1}{p{60pt}}@{}@{}*{1}{p{30pt}}@{}}
  \toprule
   \multicolumn{2}{c}{\textbf{Fault After 7 Iter.}}\\
  \cmidrule{1-2}
 ~Strategies~~~  &  ~~~~$\kappa$~~~ \\
  \cmidrule(r){1-1}\cmidrule(r){2-2}
 CCR		&	6.445		\\
 1 $\times$ Vcycle	&	1.646		\\
 2 $\times$ Vcycle	&	3.111		\\
 3 $\times$ Vcycle	&	4.445		\\
 4 $\times$ Vcycle	&	5.693		\\
 5 $\times$ Vcycle	&	6.397		\\
 \cmidrule(r){1-1}\cmidrule(r){2-2}
 1 $\times$ Wcycle	&	1.780		\\
 2 $\times$ Wcycle	&	3.404		\\
 3 $\times$ Wcycle	&	4.967		\\
 4 $\times$ Wcycle	&	6.249		\\
 5 $\times$ Wcycle	&	6.444 		\\
  \cmidrule(r){1-1}\cmidrule(r){2-2}
 1 $\times$ Fcycle	&	1.781		\\
 2 $\times$ Fcycle	&	3.405		\\
 3 $\times$ Fcycle	&	4.967		\\
 4 $\times$ Fcycle	&	6.249		\\
 5 $\times$ Fcycle	&	6.444		\\
  \cmidrule(r){1-1}\cmidrule(r){2-2}
 2 $\times$ PCG		&	0.012		\\	
 5 $\times$ PCG 	&	0.043		\\
 10 $\times$ PCG	&	0.072		\\
  \cmidrule(r){1-1}\cmidrule(r){2-2}
 10 $\times$ Smth	&	0.147		\\
 20 $\times$ Smth	&	0.236		\\
 30 $\times$ Smth	&	0.298		\\
 40 $\times$ Smth	&	0.346		\\
 50 $\times$ Smth	&	0.384		\\
 \bottomrule
  \end{tabular}
  \hspace{5pt}
   \begin{tabular}{@{}*{1}{p{60pt}}@{}@{}*{1}{p{30pt}}@{}}
  \toprule
  \multicolumn{2}{c}{\textbf{Fault After 11 Iter.}}\\
  \cmidrule{1-2}
 ~Strategies~~~  &  ~~~~$\kappa$~~~\\
  \cmidrule(r){1-1}\cmidrule(r){2-2}
  CCR 		&	10.999		\\
  1 $\times$ Vcycle	&	1.678		\\
  2 $\times$ Vcycle	&	3.225		\\
  3 $\times$ Vcycle	&	4.643		\\
  4 $\times$ Vcycle	&	5.970		\\
  5 $\times$ Vcycle	& 	7.242		\\
  \cmidrule(r){1-1}\cmidrule(r){2-2}
  1 $\times$ Wcycle	&	1.747 		\\
  2 $\times$ Wcycle	&	3.380		\\
  3 $\times$ Wcycle	&	4.951		\\
  4 $\times$ Wcycle	&	6.458		\\
  5 $\times$ Wcycle	&	7.911		\\
  \cmidrule(r){1-1}\cmidrule(r){2-2}
  1 $\times$ Fcycle	&	1.748		\\
  2 $\times$ Fcycle	&	3.381		\\
  3 $\times$ Fcycle	&	4.951		\\
  4 $\times$ Fcycle	&	6.458		\\
  5 $\times$ Fcycle	&	7.911		\\
  \cmidrule(r){1-1}\cmidrule(r){2-2}
  2 $\times$ PCG 	&	0.039		\\
  5 $\times$ PCG	&	0.104		\\
  10 $\times$ PCG 	&	0.183		\\
   \cmidrule(r){1-1}\cmidrule(r){2-2}
  10 $\times$ Smth	&	0.249		\\
  20 $\times$ Smth	&	0.387		\\
  30 $\times$ Smth	&	0.482		\\
  40 $\times$ Smth	&	0.554		\\
  50 $\times$ Smth	&	0.611		\\
   \bottomrule
 \end{tabular}
 %
 %\vspace{-0.1cm}
\end{table}

The residual error of the CCR strategy is of order $10^{-14}$ after 15 iterations such that the convergence does not saturate due to round off errors.  
We need 4.164, 6.445  or 10.999 additional V-cycles for the
case of no-local-recovery and when the failure occurs after 5, 7 or 11 iterations,
respectively, in order to achieve the same accuracy
as in the CCR strategy. This is due to the error difference, introduced by the fault,
before and after the fault which is larger the later it occurs.
These are also the maximal $\kappa$ values 
which can be achieved by a local recovery strategy. 
For the fault after 5 iterations in 
Tab. \ref{tab:DiffFaults}, we obtain similar $\kappa$ for a local recovery with 4 V-cycles,  3 W-cycles, or 3 F-cycle which is almost as good as in the CCR 
strategy.
Further, for smoothing and PCG iterations, only 
marginally small improvement can be observed in comparison to using multigrid cycles 
or the CCR strategy.
As expected %  for later faults,
for the fault after 7 iterations, 5 W- or F-cycles and for the fault after 11 iterations, more than 5 
W- or F-cycles are necessary to achieve a $\kappa$ similar to the CCR strategy. For the other strategies (smoothing and PCG iterations) no significant % ly 
improvement is obtained.
Again, we observe that W-cycles and F-cycles yield similar results, thus, F-cycles are preferred to W-cycles due to computational cost arguments.

%% file: conclusion.tex
This paper gives a first insight in constructing a fault tolerant multigrid solver. It is shown that geometric multigrid solvers are 
inherently suitable to
deal with failures of processes. The failure results in a loss of the values of a subdomain. To recover these lost values local subproblems 
with Dirichlet boundary conditions are solved by various strategies ranging from relaxation scheme, Krylov space methods to multigrid cycles.
Further, the local problems are accelerated by a {\em superman} strategy through additional parallelization of the recovery.
We introduce the reference parameter {\em Cycle Advantage} which gives the possibility to demonstrate to which extent a
recovery strategy improves the time-to-solution in terms of cycle applications in comparison to use no-recovery.
Only multigrid cycles can efficiently treat the local subproblem and even reach the same accuracy as the 
complete check-pointing recovery but with a minimized access to backup memory.

In future work, we will extend this basic idea for problems with more unknowns, incorporate the local
strategy in a global recovery strategy to balance the waiting time for all processors. 
We will enhance our considerations to problems with non-vanishing right-hand side and analyze its recovery. 
We will further develop acceleration strategies for the local recovery and include them in the HHG framework with a fault tolerant
MPI-variants.